\documentclass[11pt,a4paper]{article}
\pdfoutput=1

\usepackage{jcappub}
\usepackage[latin1]{inputenc}
\usepackage{graphicx}
\usepackage{epsfig}
\usepackage{subfigure}
\usepackage{color}
\usepackage{comment}
\usepackage{slashed}

\newcommand{\be}{\begin{equation}} 
\newcommand{\ee}{\end{equation}}
\newcommand{\bea}{\begin{eqnarray}} 
\newcommand{\eea}{\end{eqnarray}}

\newcommand{\bmp}{\noindent\begin{minipage}{16cm}}
\newcommand{\emp}{\end{minipage}\vskip 7mm} 
\def\lsim{\mathrel{\raise.3ex\hbox{$<$\kern-.75em\lower1ex\hbox{$\sim$}}}}
\def\gsim{\mathrel{\raise.3ex\hbox{$>$\kern-.75em\lower1ex\hbox{$\sim$}}}}

\newcommand{\intron}[1]{}

\subheader{\small{Preprint: HIP-2016-1/TH}}
\title{Isocurvature Constraints on Portal Couplings}

\author[a,c]{Kimmo Kainulainen,}
\author[a,c]{Sami Nurmi,}
\author[b,c]{Tommi Tenkanen,}
\author[b,c]{Kimmo Tuominen}
\author[a,c]{and Ville Vaskonen}

\affiliation[a]{Department of Physics, University of Jyv{\"a}skyl{\"a}, \\
                      P.O.Box 35 (YFL), FI-40014 University of Jyv{\"a}skyl{\"a}, Finland}
\affiliation[b]{Department of Physics, University of Helsinki \\
                      P.O.~Box 64, FI-00014, Helsinki, Finland}
\affiliation[c]{Helsinki Institute of Physics, \\
                      P.O.~Box 64, FI-00014, Helsinki, Finland}   
                            
\emailAdd{kimmo.kainulainen@jyu.fi}
\emailAdd{sami.t.nurmi@jyu.fi}
\emailAdd{tommi.tenkanen@helsinki.fi}
\emailAdd{kimmo.i.tuominen@helsinki.fi}
\emailAdd{ville.vaskonen@jyu.fi}

\abstract{We consider portal models which are ultraweakly coupled with the Standard Model, and confront them with observational constraints on dark matter abundance and isocurvature perturbations. We assume the hidden sector to contain a real singlet scalar $s$ and a sterile neutrino $\psi$ coupled to $s$ via a pseudoscalar Yukawa term. During inflation, a primordial condensate consisting of the singlet scalar $s$ is generated, and its contribution to the isocurvature perturbations is imprinted onto the dark matter abundance. We compute the total dark matter abundance including the contributions from condensate decay and nonthermal production from the Standard Model sector. We then use the Planck limit on isocurvature perturbations to derive a novel constraint connecting dark matter mass and the singlet self coupling with the scale of inflation: $m_{\rm DM}/{\rm GeV}\lsim 0.2\lambda_{\rm s}^{\scriptscriptstyle 3/8} \left(H_*/10^{\scriptscriptstyle 11}{\rm GeV}\right)^{\scriptscriptstyle -3/2}$. This constraint is relevant in most portal models ultraweakly coupled with the Standard Model and containing light singlet scalar fields.}


\begin{document}
\maketitle
%

%
\section{Introduction}
%

After the detection of Higgs boson with $m_{\rm h} = 125$ GeV there has been considerable interest towards its cosmological ramifications. During inflation the Standard Model (SM) Higgs is an energetically subdominant light field\footnote{Here we do not consider the possibility of Higgs inflation which assumes a UV fixed point with a large non-minimal Higgs-curvature coupling \cite{Bezrukov:2007ep}.}, which acquires nearly scale invariant fluctuations \cite{Espinosa:2007qp,EliasMiro:2011aa,Enqvist:2013kaa}.  Long wave-length fluctuations tend to displace the field from its vacuum locally, generating an effective primordial Higgs condensate over the observable universe~\cite{Starobinsky:1994bd}. Similar condensates are formed for all light spectator scalars frequently encountered in SM extensions. Primordial condensates set specific non-equilibrium initial conditions for the early universe. They can leave direct observational imprints ranging from non-thermal dark matter production~\cite{Nurmi:2015ema} to the generation of baryon asymmetry~\cite{Enqvist:2014zqa,Kusenko:2014lra}. A consistent treatment of condensates is therefore an integral part of testing SM extensions, in addition to imposing other direct observational constraints. The requirement of electroweak vacuum stability against inflationary fluctuations  \cite{Espinosa:2007qp,EliasMiro:2011aa} provides another novel window to test spectator couplings. Recently the stability conditions have been extensively studied \cite{Gonderinger:2009jp, EliasMiro:2012ay,Kobakhidze:2013tn,Enqvist:2013kaa, Fairbairn:2014zia, Hook:2014uia, Espinosa:2015qea} accounting both for the non-minimal curvature coupling  ~\cite{Herranen:2014cua, Herranen:2015ima,Moss:2015gua} as well as eventual couplings to new physics ~\cite{Chao:2012mx,Blum:2015rpa}.

In this work we investigate in detail cosmological constraints on Higgs portal extensions of SM. We will in particular analyse the effects of the inflationary initial conditions. Many of the cosmologically interesting features of portal scenarios are captured by the simplest effective model featuring a scalar singlet coupled to Higgs via $\lambda_{\rm hs} s^2 \Phi^{\dagger}\Phi$~\cite{McDonald:1993ex,Espinosa:2011ax,Cline:2013gha,Cline:2012hg}. Here we include also a singlet fermion field, i.e. a sterile neutrino, in the portal sector, with a Yukawa coupling to the singlet scalar~\cite{LopezHonorez:2012kv,Fairbairn:2013uta,Alanne:2014bra,Esch:2014jpa,Kainulainen:2015cde}. In this setup both the singlet scalar and the singlet fermion can be dark matter. If the portal coupling is very weak the portal sector will never thermalize with the SM fields. In that case, dark matter can be produced by two mechanisms: by decay of the primordial singlet condensate and by non-equilibrium decays of SM particles. The latter is often referred to as the freeze-in~\cite{McDonald:2001vt,Hall:2009bx, Yaguna:2011qn, Blennow:2013jba,Elahi:2014fsa, Kang:2015aqa, Merle:2013wta, Klasen:2013ypa, Adulpravitchai:2014xna, Merle:2014xpa, Merle:2015oja} as opposed to the usual freeze-out of thermal relics.  As shown in~\cite{Nurmi:2015ema} the decay of the primordial condensate can easily dominate the dark matter yield. 

A primordial condensate originating from inflationary fluctuations in general contains an isocurvature component uncorrelated with the SM sector. When dark matter is sourced by such a condensate, an isocurvature fluctuation gets imprinted to it, and if the dark sector is weakly coupled to SM, this isocurvature component persists. However, isocurvature is heavily constrained by observations of the Cosmic Microwave Background (CMB). This leads to stringent bounds on model parameters in portal scenarios. In our specific model we find a novel constraint connecting the isocurvature dark matter mass and the singlet self coupling to the inflationary scale, $m_{\rm DM}/{\rm GeV}\lsim 0.2\lambda_{\rm s}^{\scriptscriptstyle 3/8} \left(H_*/10^{\scriptscriptstyle 11}{\rm GeV}\right)^{\scriptscriptstyle -3/2}$, 
given that the portal coupling is small enough, $|\lambda_{\rm hs}| \lsim 10^{-7}$. This bound connects in intriguing way the standard electroweak scale physics (freeze-in production of dark matter) to the isocurvature constraints sensitive to physics up to the inflationary scale. It should be stressed that bounds like this are generic to most weakly coupled portal DM models with light scalar fields.

The paper is organised as follows. In Section \ref{model}, we present the model and discuss inflationary initial conditions and isocurvature perturbations. In Section \ref{productionHighT} we identify the dark matter candidates in the setup and investigate the different channels for dark matter production. In Section \ref{results} we contrast the dark matter abundance and isocurvature perturbations against the observational data and present the imposed constraints on couplings and mass scales. In Section \ref{conclusions} we summarise and present our conclusions.

%
\section{The Model}
\label{model}
%

We consider Higgs portal extensions of the Standard Model, where the portal sector includes a real singlet (pseudo)scalar $s$ and a fermion $\psi$. We assume that the Lagrangian is invariant under the parity transformation $\psi(t,x)\to\gamma^0\psi(t,-x)$, $s(t,x) \to -s(t,-x)$. 
The fermionic part of the portal sector is
\be
\label{Lpsi}
\mathcal{L}_\psi = \bar\psi(i\slashed\partial - m_\psi)\psi + igs\bar{\psi}\gamma_5\psi~,
\ee
and the most general renormalizable scalar potential is given by
\be
V(\Phi,s) = \mu_{\rm h}^2\Phi^\dagger\Phi+\lambda_{\rm h}(\Phi^\dagger\Phi)^2+\frac{\mu_{\rm s}^2}{2} s^2+\frac{\lambda_{\rm s}}{4}s^4+\frac{\lambda_{\rm hs}}{2}\Phi^\dagger\Phi s^2~.
\label{potential}
\ee
Here $\Phi$ is the SM Higgs doublet with the standard kinetic terms. In the unitary gauge the Higgs doublet is written as $\sqrt{2}\Phi^{\rm T}=(0,v+h)$, where $v = 246~\mathrm{GeV}$ at $T=0$. We assume throughout the paper that $\mu_{\rm s}^2>0$ and $m_{\rm s}^2 \equiv \mu_{\rm s}^2+\lambda_{\rm hs} v^2/2 > 0$, so that the minimum of the potential is at $s=0$ and $m_{\rm s}$ is the physical mass  of $s$ in zero temperature vacuum. These imply an upper limit on the portal coupling, $\lambda_{\rm hs} < 2m_{\rm s}^2/v^2$. We also assume that $\lambda_{\rm h}>0$, $\lambda_{\rm s}>0$ and $\lambda_{\rm hs} > - 2\sqrt{\lambda_{\rm h} \lambda_{\rm s}}$ guaranteeing that the tree level potential is bounded from below.

%
\subsection{Initial values of the scalar fields}
\label{inflationary_constraints}
%

For field values sufficiently below the Planck scale,  both the Higgs and the scalar singlet are light during inflation, $V'' \ll H^2$, and acquire nearly scale invariant fluctuations  \cite{Espinosa:2007qp, Enqvist:2014zqa}. Although $\langle h\rangle = \langle s \rangle = 0 $ over the entire inflating patch, the accumulation of long-wavelength fluctuations generates effective Higgs and singlet condensates over the observable universe. If inflation lasts longer than the minimal $N\sim 60$ e-folds, the spatial averages computed over the observable patch will in general significantly differ from zero. A quantitative estimate for the observable average is given by the root mean square $\phi_{*}\equiv \sqrt{\langle \phi^2 \rangle}$, $\phi = h,s$, of fluctuations over the entire inflating patch \cite{Starobinsky:1994bd} 
\be
\label{h,s_*}
h_{*}\simeq 0.36 \frac{H_*}{\lambda_{\rm h}^{\scriptscriptstyle 1/4}}\ ,\qquad s_{*}\simeq 0.36\frac{H_*}{\lambda_{\rm s}^{\scriptscriptstyle 1/4}}\ .
\ee
Here $H_*$ is the value of the Hubble parameter at the horizon crossing of observable modes. Here we have assumed that the portal coupling is small $|\lambda_{\rm hs}| \ll \sqrt{\lambda_{\rm h}\lambda_{\rm s}}$. For $ 0.01 \lsim \lambda_{\rm \phi} \lsim 1$, this yields a slight overestimate as the scalars become effectively massive before reaching the value $\phi_*$. However, we have checked that the correction is at most an order of magnitude even when $\lambda_{\rm \phi} \sim 1$. 

In Eqs. (\ref{h,s_*}) we have implicitly assumed that the non-minimal curvature couplings $\xi_{\rm h} R h^2$ and  $\xi_{\rm s} R s^2$ (necessarily generated by radiative corrections) are negligible: $| \xi_{\rm h,s}|\ll 1$. For the Higgs field this is consistent with electroweak vacuum stability only for $H_{*} \lsim 10^{11}$ GeV  \cite{Herranen:2014cua}.  For $H_{*} \gtrsim 10^{11}$ GeV the stability requires $\xi_h\gtrsim 0.1$. This renders the Higgs effectively massive and no condensate is formed.
On the other hand, the stability is not affected by the weakly coupled singlet and we may take $|\xi_{\rm s}|\ll 1$ irrespectively of the inflationary scale. A singlet condensate is then necessarily formed and we will take~\eqref{h,s_*} as the initial condition for $s$ after inflation. For the Higgs field we take~\eqref{h,s_*} as the initial condition when $H_*\lsim 10^{11}$ GeV, whereas for higher inflationary scales we use $h_*=0$. We assume instantaneous reheating after the end of inflation, where a thermal bath of SM particles with a temperature $T_{\rm reh} \sim (H_*M_P)^{1/2}$ is produced. After this the temperature falls as $T\propto a^{-1}$. The Higgs condensate thermalises around $T=10^{-2}T_{\rm reh}$~\cite{Enqvist:2014zqa} and has no impact on the dark matter yield (see \cite{Enqvist:2015sua,Figueroa:2015rqa,Enqvist:2014tta} for its evolution in $T=0$ background). The singlet condensate does not feel the thermal bath if $|\lambda_{\rm hs}| \lsim 10^{-7} $ and, consequently, in this regime a sizeable fraction of dark matter in the portal sector will be produced out of the primordial condensate~\cite{Nurmi:2015ema}.

%
\subsection{Vacuum stability}
%

Due to the very weak portal coupling $|\lambda_{\rm hs}|\lsim 10^{-7}$, the analysis of the SM vacuum stability during inflation is not affected by the new scalar and fermion fields. However, the $s$-dependent part of the scalar potential may develop another minimum at a nonzero field value as the fermion coupling in Eq.~\eqref{Lpsi} gives a negative contribution to the beta function of $\lambda_{\rm s}$. The energy density at such symmetry breaking vacuum could, in extreme case, be very large and negative, and potentially lead to problems such as collapsing universe, see e.g. \cite{Hook:2014uia, Espinosa:2015qea}. Even though such pathologies are not expected for all vacua with nonzero $s$, we choose to restrict ourselves in the regime where the $s=0$ vacuum is stable against inflationary fluctuations. 

\begin{figure}
\begin{center}
\includegraphics[width=.46\textwidth]{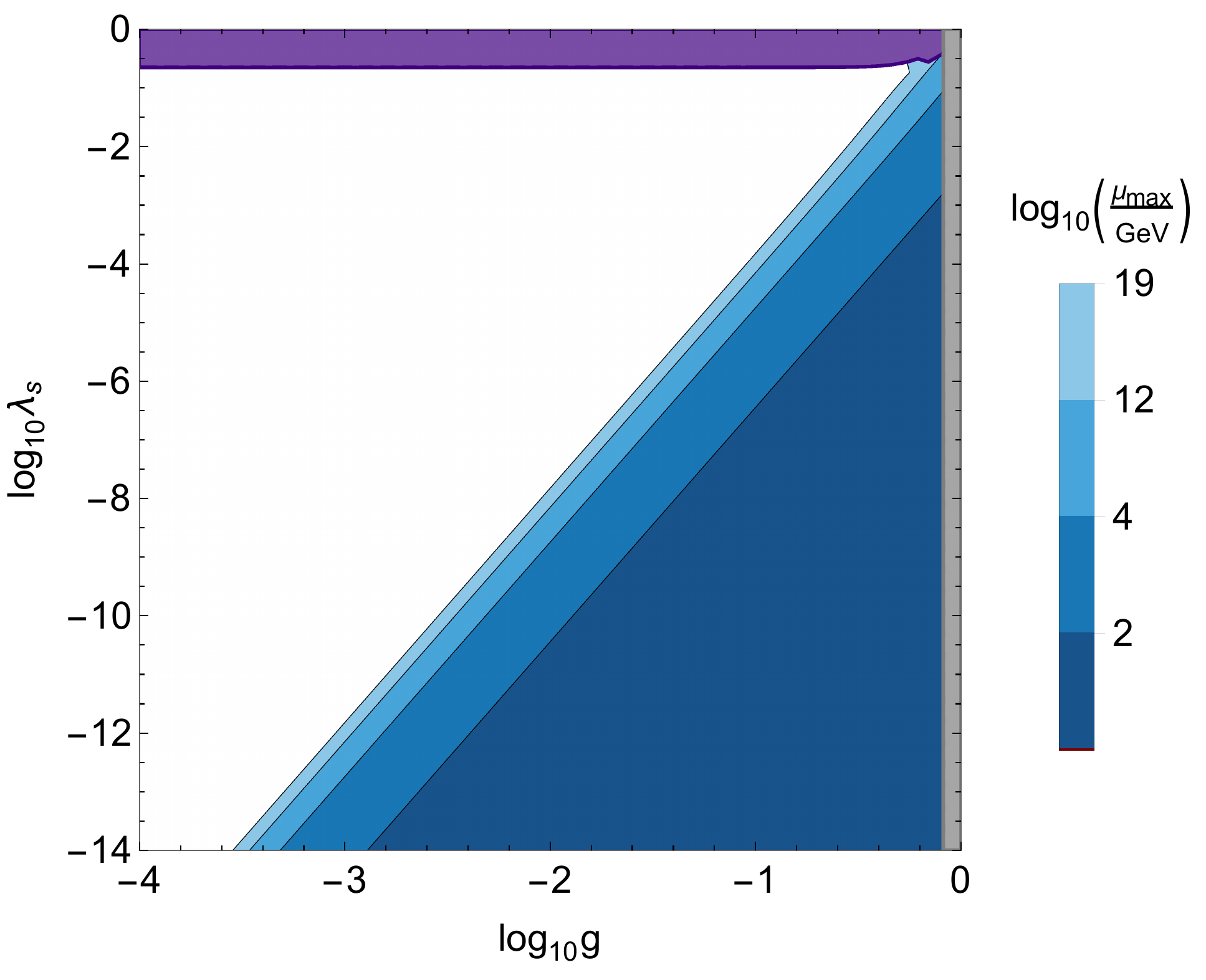}
\caption{In the blue regions $\lambda_{\rm s}$ becomes negative at the scale $\mu_{\rm max}$ as indicated by the bar to the right of the plot. The axes show the values of $\lambda_{\rm s}$ and $g$ at scale $\mu = M_Z$. In the purple (gray) region $\lambda_{\rm s}$  ($g$) has a Landau pole below the Planck scale.}
\label{fig1}
\end{center}
\end{figure}

Neglecting contributions induced by Higgs loops, which are heavily suppressed by the small coupling $\lambda_{\rm hs}$, the beta functions for $\lambda_{\rm s}$ and $g$ are
\be
16\pi^2 \beta_{\lambda_{\rm s}} = 18\lambda_{\rm s}^2 - 6 g^4, \quad 16\pi^2 \beta_g = 3 g^3.
\label{betafunctions}
\ee
Let us denote by $\mu_{\rm{max}}$ the scale above which the coupling $\lambda_{\rm s}$ becomes negative. Dimensionally, the height of the barrier between the $s=0$ vacuum and the possible negative energy vacuum is \nolinebreak[4] $V_{\rm max}^{\scriptscriptstyle 1/4}\sim \mu_{\rm max}$.  Inflationary fluctuations may push the field $s$ to the negative energy vacuum if $H_\ast\gsim V_{\rm max}^{\scriptscriptstyle 1/4}$. Avoiding this imposes a constraint between $g$ and $\lambda_{\rm s}$ shown in \nolinebreak[4] Fig.~\ref{fig1}. Specifically, if we take $\mu_{\rm max}=10^{12}$ GeV, we need to have $g\lsim1.5\lambda_{\rm s}^{\scriptscriptstyle 1/4}$. We have also computed the running of couplings from Eq.~(\ref{betafunctions}) to determine the region of the parameter space where the coupling $\lambda_{\rm s}$ has a Landau pole below $H_*$.

%
\subsection{Isocurvature perturbations}
\label{isocurvature_perturbations}
%

For $|\lambda_{\rm hs}|\lsim 10^{-7}$ the portal sector will never thermalise with the SM fields. The dark matter component sourced by the primordial singlet condensate will therefore retain its isocurvature  fluctuation, whose amplitude is heavily constrained by CMB observations. This constrains the energy density of the singlet condensate to be small. Consequently, the isocurvature perturbations have only a negligible effect on the evolution of adiabatic perturbations. Hence, as the singlet fluctuations are assumed to be decoupled from the inflaton sector, the adiabatic and isocurvature perturbations can be treated as totally uncorrelated.  

Here we assume that both the scalar and fermion masses are above the electronvolt scale $m_{s,\psi}\gsim 1$ eV.  In this regime we find that they constitute non-relativistic matter at the time of CMB formation. The isocurvature perturbation on superhorizon scales is then given \nolinebreak[4] by
\be
S= \left(\frac{\delta (\rho_{\rm s_0} + \rho_{\rm CDM})} {\rho_{\rm s_0} + \rho_{\rm CDM}}-\frac{3}{4}\frac{\delta\rho_{\gamma}}{\rho_{\gamma}} \right) = \frac{\rho_{\rm s_0}}{\rho_{\rm s_0} + \rho_{\rm CDM}}\left(-3\zeta+\frac{\delta\rho_{\rm s_0}}{\rho_{\rm s_0}}\right) \,.
\ee
Here $\rho_{\gamma}$ is the radiation energy density, $\rho_{\rm s_0}$ denotes the non-relativistic matter sourced by the singlet condensate and $\rho_{\rm CDM}$ denotes all other non-relativistic matter which we assume to be adiabatic: $\delta\rho_{\rm CDM}/\rho_{\rm CDM} = (3/4) {\delta\rho_{\gamma}}/{\rho_{\gamma}} = -3 \zeta$. All perturbations are evaluated at  photon decoupling $T_{\rm dec}\sim 0.3$ eV. Using $\langle\zeta\delta\rho_{\rm s_0}\rangle=0$ and denoting $\delta_{\rm s_0} = \delta \rho_{\rm s_0}/\rho_{\rm s_0}$, the spectrum of isocurvature perturbations becomes 
\be
\label{P_S}
{\cal P}_{S} =  \left(\frac{\rho_{\rm s_0}}{\rho_{\rm s_0} + \rho_{\rm CDM}}\right)^2\left(9+\frac{{\cal P}_{\delta_{\rm s_0}}}{{\cal P}_{\zeta}}\right){\cal P}_{\zeta} \equiv \frac{\beta}{1-\beta} {\cal P}_{\zeta}\ .
\ee
This defines the isocurvature parameter $\beta$. Planck data constrains the isocurvature  contribution to be at most a few percents $\beta\lsim 0.05$~\cite{Ade:2015lrj} for isocurvature modes uncorrelated with adiabatic perturbations.

We need to estimate the quantity ${\cal P}_{\delta_{\rm s_0}}$ at the time of CMB formation. Here the field $s$ oscillates in effectively quadratic potential and the envelope of the field is related to the inflationary initial condition through  $s \propto s_*^{\scriptscriptstyle 3/4}$~\cite{Enqvist:2009zf}. Correspondingly, we get $\delta_{\rm s_0} = 2\delta s/ s = (3/2) \delta s_{*}/ s_{*}$. The inflationary fluctuations $\delta s_*$ around the effective background field value $s_*$ have the usual spectrum of a massless scalar, ${\cal P}_{\delta s_*} = (H_*/2\pi)^2$. Thus, we get the result 
\be
\label{P_delta}
{\cal P}_{\delta_{\rm s_0}}  = \frac{9}{4}\frac{{\cal P}_{\delta s_*}}{s_{*}^2} \simeq 0.4 \lambda_{\rm s}^{1/2} ,
\ee
where we have used Eq.~\eqref{h,s_*} for $s_*$ in the last step. Finally, using ${\cal P}_{\zeta}\simeq 2.2 \times 10^{-9}$  and $\beta\lsim 0.05$~\cite{Ade:2015lrj} in Eqs.~\eqref{P_S} and~\eqref{P_delta}, we get an absolute upper bound for today's dark matter abundance sourced by the singlet condensate:
\be
\label{isocurvature_constraint}
\frac{\Omega_{\rm DM}^{({\rm s}_0)} h^2}{0.12} \lsim 1.6\times 10^{-5}\lambda_{\rm s}^{-1/4} .
\ee
This absolute upper bound provides stringent constraints on the model parameters, as we will discuss in Sections \ref{productionHighT} and \ref{results}.

%
\section{Dark Matter production}
\label{productionHighT}
%

In the model discussed here both the singlet fermion $\psi$ and the singlet scalar $s$ can be dark matter particles. Singlet scalar particles are generated through the decay of the primordial singlet condensate~\eqref{h,s_*} and through the standard freeze-in mechanism dominated by decays of Higgs particles at the electroweak scale~\cite{McDonald:2001vt,Hall:2009bx, Yaguna:2011qn, Blennow:2013jba,Elahi:2014fsa, Kang:2015aqa, Nurmi:2015ema, Merle:2013wta, Klasen:2013ypa, Adulpravitchai:2014xna, Merle:2014xpa,Merle:2015oja}. On the other hand, singlet fermions can be generated through decays of either the singlet scalar condensate or singlet scalar particles through the Yukawa coupling~\eqref{Lpsi}. 

%
\subsection{Particle production from a primordial field}
\label{bgDynamics}
%

The singlet condensate does not feel the thermal bath of SM particles when $|\lambda_{\rm hs}|\lsim 10^{-7}$. After it becomes massive and starts to oscillate, its energy density is diluted by the expansion of space. The oscillating field can decay to singlet scalar particles and singlet fermions and to Higgs bosons. The evolution of the energy density of the condensate is determined by
\be
\label{condensate_Boltzmann}
\dot{\rho}_{\rm s_0} + 3H (1+w) \rho_{\rm s_0} = -  \left( \left\langle\Gamma_{s_0\rightarrow ss} \right\rangle+ \left\langle\Gamma_{s_0\rightarrow \bar{\psi}\psi}\right\rangle \right) \rho_{\rm s_0}\ ,
\ee
where $\left\langle \Gamma_{i}\right\rangle$ are the decay rates of the condensate averaged over one oscillation. The parameter $w=1/3$ if the condensate oscillates in the {\em quartic} regime (where potential is dominated by the quartic term) and $w=0$ if the condensate oscillates in the {\em quadratic} regime. In quartic regime both $s_0\rightarrow ss$ and $s_0\rightarrow \bar{\psi}\psi$ channels are open, but the latter is not effective due to Fermi-blocking (see appendix \ref{fermi_statistics} for details), whereas in the quadratic regime the channel $s_0\rightarrow ss$ is kinematically blocked. The effect of other channels, such as $s_0\rightarrow hh$ and $s_0\rightarrow sss$ and of condensate induced processes $h \rightarrow sh$, $s \rightarrow ss$ and $s \rightarrow hh$ are negligible. All relevant decay rates are given in Appendix \ref{append_decay_rates}.

Decay processes have negligible effect on the condensate dynamics until their rate becomes comparable to the Hubble scale; up to this point evolution is affected only by the expansion of space. The decay channel whose rate $\left\langle \Gamma_{i}\right\rangle$ becomes equal to the Hubble scale $H$ first will dominate and defines the condensate decay temperature via $\left\langle \Gamma_{i}\right\rangle(T_{\rm dec}) = H(T_{\rm dec})$. To estimate the dark matter abundance sourced by the condensate we need to identify the dominant channel and determine how the fermion and scalar masses are ordered. We find that in the present case the abundance of dark matter produced out of the condensate can be expressed in the following general form:
\be
\label{highT_abundance}
\begin{aligned}
\frac{\Omega_{\rm DM}^{({\rm s}_0)} h^2}{0.12} &\simeq \frac{1}{0.12} \frac{h^2}{\rho_c} m_{\rm DM} \frac{\rho_{\rm s_0}(T_{\rm dec})}{E_{\rm DM}} \frac{g_{{\rm s}*}(T_0)}{g_{{\rm s}*}(T_{\rm dec})} \left(\frac{T_0}{T_{\rm dec}}\right)^3 \,. \\
& \simeq 
C \times 10^{-5} \left( \frac{\omega_{\rm dec}}{E_{\rm DM}} \right) \lambda_{\rm s}^{-5/8} \left(\frac{m_{\rm DM}}{\mathrm{GeV}}\right) \left(\frac{H_*}{10^{11}\mathrm{GeV}}\right)^{3/2} \,.
\end{aligned}
\ee
Here $\rho_c$ is the critical density and $T_0$ is the CMB temperature today, $g_{{\rm s}*}$ is the effective number of entropy degrees of freedom. $E_{\rm DM}$ is the average energy of the dark matter particles produced in decays and $\omega_{\rm dec}$ is defined as the oscillation frequency of the lowest Fourier mode of the $s_0$ field during its decay. Finally the constant $C \approx 7.5$ in the quadratic and $C \approx 6.4$ in the quartic regime (see appendix~\ref{append_decay_rates} for details). 

The energy $E_{\rm{DM}}$ and the mass $m_{\rm{DM}}$ may have the following combinations of values: First, if $s$ is stable (against the decay $s\rightarrow \psi\bar\psi$) above the CMB-formation temperature, and the condensate decays through the channel $s_0\rightarrow ss$, then $m_{\rm DM}=m_{\rm s}$. Since this decay is controlled by operator $s_0^2ss$ and the frequency of $s_0^2$ at decay temperature $2\omega_{\rm dec}$, we may approximate $E_{\rm DM}\approx \omega_{\rm dec}$. In all other cases $m_{\rm DM}=m_\psi$ irrespective of the mass hierarchy, even when condensate first decays to, now unstable $s$, because the inverse decay rate $\psi\bar\psi \rightarrow s$ is always negligibly small. Also $E_{\rm DM}\approx \omega_{\rm dec}/2$ in all these cases. If condensate decays directly to fermions, this is because decay is controlled by operator $s_0\psi\bar\psi$, and the frequency of $s_0$ is $\omega_{\rm dec}$. When condensate decays first to unstable scalars we get the same result because each primary $s$-particle now has energy $\omega_{\rm dec}$, but as they decay each daughter fermion gets half the energy of the parent state. Finally, there is the possibility that the condensate does not decay by the time of recombination. In this case the condensate itself plays the role of the dark matter and we find
\be
\frac{\Omega_{\rm DM}^{({\rm s}_0)} h^2}{0.12} = \frac{1}{0.12} \frac{h^2}{\rho_c} \rho_{\rm s_0}(T_0) \approx 6.4 \times 10^{-5}\lambda_{\rm s}^{-5/8} \left(\frac{m_{\rm s}}{\mathrm{GeV}}\right) \left(\frac{H_*}{10^{11}\mathrm{GeV}}\right)^{3/2} .
\ee

\begin{figure}
\begin{center}
\includegraphics[width=0.78\textwidth]{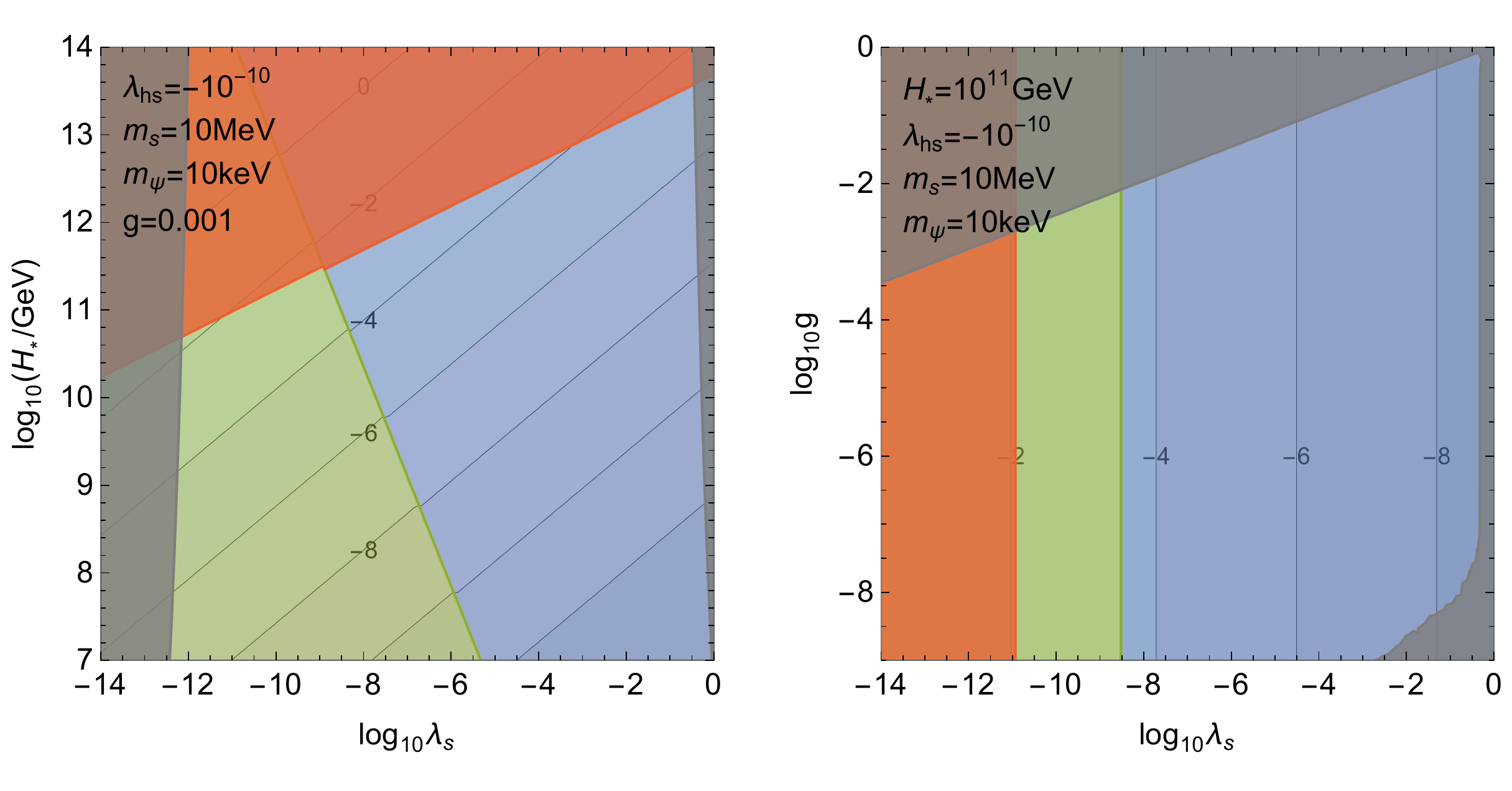}
\caption{In the blue region the oscillating $s_0$ field decays to scalars in the quartic regime. In the green region $s_0$ decays to fermions in the quadratic regime. The red region is excluded by the isocurvature constraint. In the grey region $V_{\rm max}^{\scriptscriptstyle 1/4} > H_*$, or $\lambda_{\rm s}$ has a Landau pole below the scale $H_*$. Contours show the logarithm of the dark matter abundance, $\log_{10}(\Omega_{\rm DM}^{\scriptscriptstyle ({\rm s}_0)} h^2/0.12)$, produced from the scalar condensate.}
\label{fig2}
\end{center}
\end{figure}

In Fig.~\ref{fig2} we display results from a calculation that accounts for all the details described above. We show regions in self-coupling $\lambda_{\rm s}$ and either the inflationary scale $H_*$ (left panel) or the fermion coupling $g$ (right panel), where the condensate decays either to fermions (green areas) or scalars (blue areas). We show also regions (red areas) excluded by isocurvature constraint as well as contours of dark matter abundance produced from the condensate decay. Gray regions show where our calculation is not self-consistent: either $\lambda_{\rm s}$ has a Landau pole below the inflationary scale, or the $s=0$ vacuum is not stable against inflationary fluctuations. It should be noted that the isocurvature bound~\eqref{isocurvature_constraint} is much tighter than the limit where dark matter would overclose the universe.

%
\subsection{Dark matter production via freeze-in}
\label{totalDM}
%

In addition to the dark matter production from the primordial field considered above, also  direct production from Standard Model needs be taken into account. For weakly coupled portal models the standard thermal freeze-out mechanism is inefficient, but nonthermal {\em freeze-in} production~\cite{Hall:2009bx} can easily be efficient at temperatures below the EW scale.
In freeze-in scenarios dark matter particles never reach thermal equilibrium with the SM particles, and the initial occupation number of the dark matter particles are either zero, or negligibly small. The dark matter production is always dominated by the Higgs boson decay channel $h\rightarrow ss$. In this case the dark matter abundance produced by freeze-in mechanism is
\be
\frac{\Omega_{\rm DM}^{({\rm fi})} h^2}{0.12} = 5.3\times 10^{21} N \lambda_{\rm hs}^2 \left(\frac{m_{\rm DM}}{\rm GeV}\right).
\label{eq:dmfromsm}
\ee
If dark matter is the singlet scalar particle ($m_{\rm s} < 2m_\psi$) then $m_{\rm{DM}}=m_{\rm s}$ and $N=1$. If dark matter is the singlet fermion ($m_{\rm s} > 2m_\psi$) then $m_{\rm{DM}}=m_\psi$ and the factor $N=2$ accounts for the fact that two fermions are produced in each subsequent $s$ decay. We can now simply sum up the yields from the scalar condensate and Higgs decay to obtain
\be
\label{lowT+highT_results}
\Omega_{\rm DM} = \Omega_{\rm DM}^{({\rm s}_0)} + \Omega_{\rm DM}^{({\rm fi})}.
\ee
This assumed that the dark matter component created earlier by the primordial field may be neglected when computing the yield~(\ref{eq:dmfromsm}). This is a self-consistent condition for a sufficiently heavy, weakly coupled dark matter satisfying the overclosure constraint in general, and here in particular in regions allowed by the much more stringent isocurvature constraint. Indeed, when $\psi$ is the dark matter particle, its mass is bounded from below by the Tremaine-Gunn limit~\cite{Tremaine:1979we,Boyarsky:2008ju} $m_{\rm \psi} \gsim 0.1$ keV. 

%
\section{Results}
\label{results}
%

We now move on to present results for the dark matter yield accounting for all the decay channels of the primordial singlet condensate and the freeze-in production through Standard Model decays.   

\begin{figure}
\begin{center}
\includegraphics[height=0.78\textwidth]{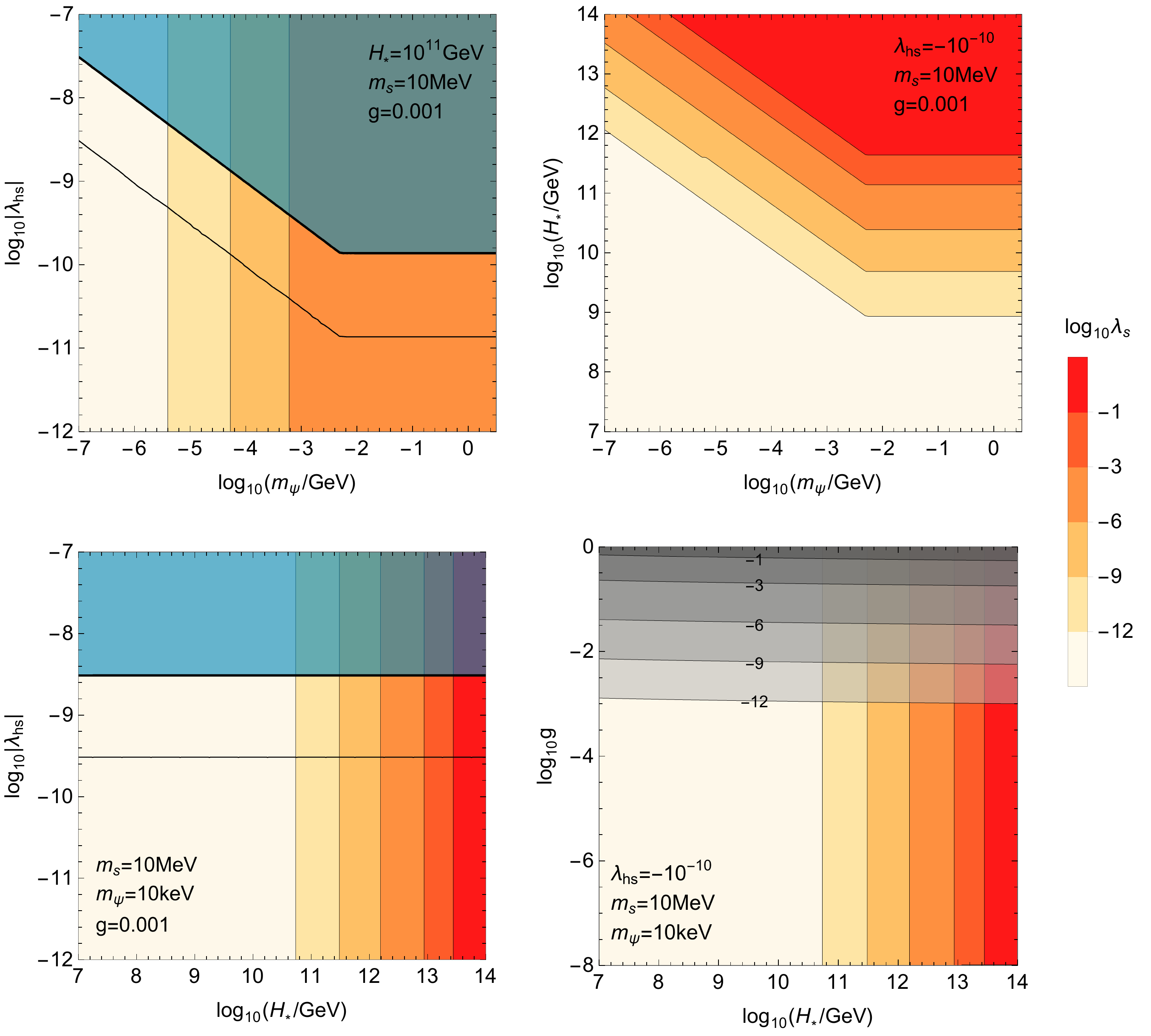}
\caption{Isocurvature constraint excludes areas right from the contours labelled by different values of $\lambda_{\rm s}$ (as indicated by the colour bar). In blue regions in the left panels the total dark matter abundance is too large, $\Omega_\mathrm{DM} h^2>0.12$. Thin line shows the contour with $\Omega_\mathrm{DM} h^2/0.12 = 0.01$. In the bottom right plot the grey regions show where $V_{\rm max}^{\scriptscriptstyle 1/4}<H_*$ for different values of $\log_{10}\lambda_{\rm s}$ indicated in figure. Note that the portal coupling $\lambda_{\rm hs}$ is negative in all figures.}
\label{fig3}
\end{center}
\end{figure}

A particularly important consequence of a primordial scalar background in the weakly coupled portal sector is the generation of isocurvature perturbations.  In \cite{Nurmi:2015ema} it was argued that the isocurvature bounds could be alleviated if the primordial condensate decays at least partially into relativistic degrees of freedom. Here we have investigated this effect including singlet fermions in the portal sector and treating their mass as a fee parameter. As can be seen in the upper right panel of Fig.~\ref{fig3}, the lower limit on the singlet self-coupling $\lambda_{\rm s}$ indeed gets relaxed as the fermion mass decreases. However, the allowed parameter space is still strongly constrained indicating the very general role of isocurvature bounds in weakly coupled portal extensions of the Standard Model. 

By making use of the isocurvature constraint~\eqref{isocurvature_constraint} and total abundance of particles produced out from a primordial scalar field~\eqref{highT_abundance}, we can derive a rough upper bound on the mass of the particle constitituting an isocurvature component of the dark matter,
\be
\label{isocurvatureConstrainedMass}
\left(\frac{m_{\rm DM}}{\rm GeV}\right)\lsim 0.2 \lambda_{\rm s}^{3/8} \left(\frac{H_{*}}{10^{11}{\rm GeV}}\right)^{-3/2}.
\ee
We stress that the formation of primordial condensates is a typical consequence in a theory which contains  scalar fields. Therefore we expect that qualitatively similar results would constrain the masses and couplings also in other, more generic portal models and extensions of the SM.

The total dark matter abundance together with the isocurvature constraint are shown in Fig.~\ref{fig3}. Blue regions in left panels are excluded because a too large portal coupling $|\lambda_{\rm hs}|$ leads to an overproduction of dark matter from Higgs boson decay. The red regions are excluded by the isocurvature constraint the bound being stronger (weaker) for smaller (larger) values of the self-coupling $\lambda_{\rm s}$. In the grey areas in the right bottom panel,  corresponding to large fermion coupling, our calculation is not self-consistent. Overall, the isocurvature bound constrains the singlet self-coupling $\lambda_{\rm s}$ from below and the bound gets tighter as the inflationary scale $H_*$ increases. The portal coupling $|\lambda_{\rm hs}|$ on the other hand is constrained from above by the abundance of adiabatic dark matter which should not exceed the observed value. 

\begin{figure}
\begin{center}
\includegraphics[width=0.78\textwidth]{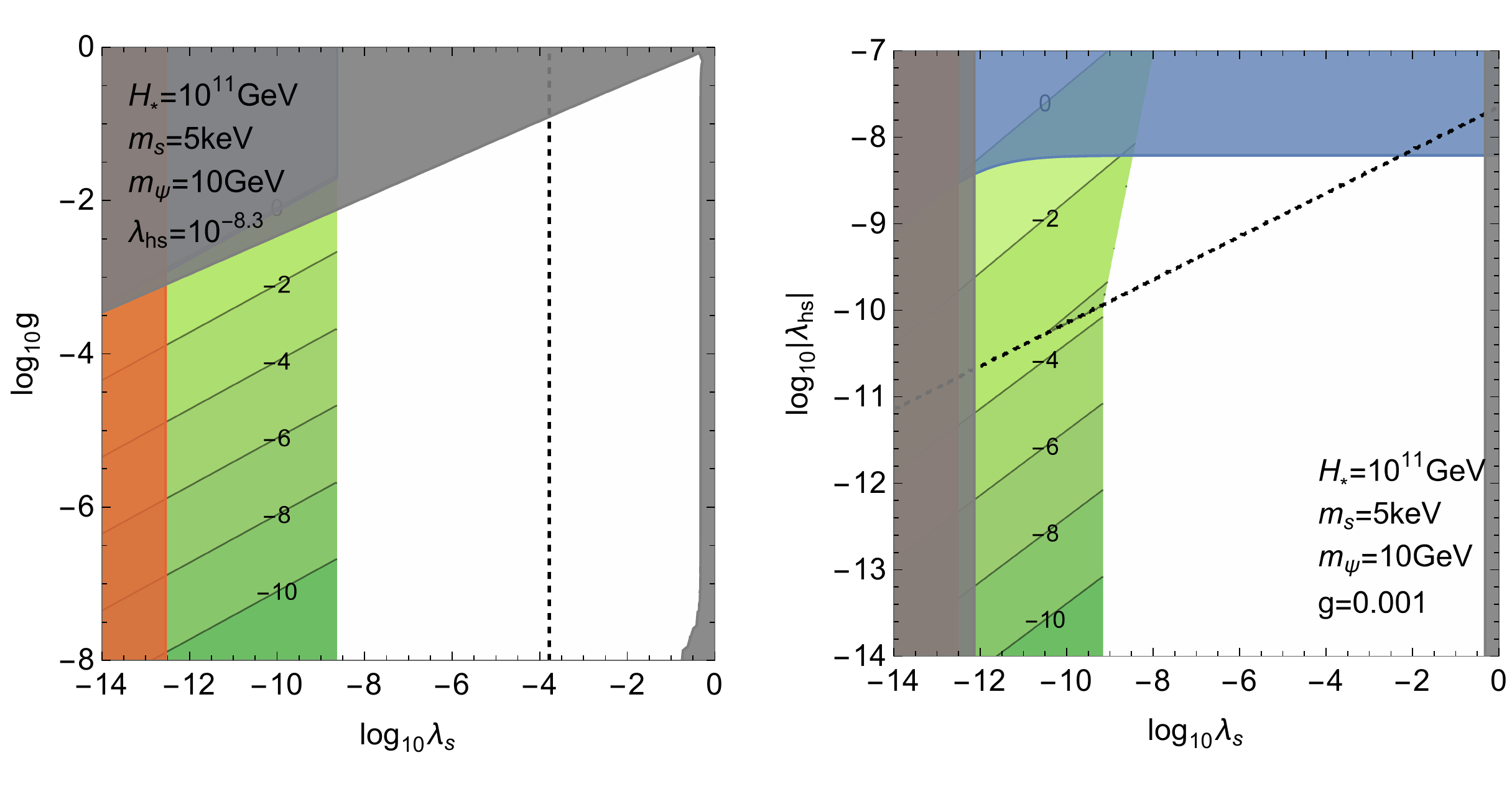}
\caption{Contours show the abundance of singlet fermions, $\log_{10}(\Omega_\psi h^2/0.12)$, produced by Higgs decay, $H\to\psi\bar\psi$, via condensate induced mixing of $s$ and $h$. Red and grey areas are as in figure~\ref{fig2}. In the blue region the total dark matter abundance is too large ($\Omega_{\rm DM} h^2>0.12$). In the left panel, in the area left from the dashed line (in the right panel, in the area above the dashed line) the transition to the quadratic regime happens at $T>T_{\rm EW}$. In the white region the primordial field has decayed by $T=T_{\rm EW}$.}
\label{fig4}
\end{center}
\end{figure}

For a very small self-coupling $\lambda_{\rm s}$ (see for example Fig.~\ref{fig4}) the condensate does not decay before the electroweak phase transition at $T=T_{\rm EW}$. After the transition a nonzero Higgs vev induces a mixing between $h$ and $s$, which gives a new channel for $\psi$ production. We have checked that the production of the singlet fermions from Higgs boson decay via the mixing is negligible in most of the allowed region, however. Results of a calculation including both condensate and freeze-in production of dark matter are shown in Fig.~\ref{fig4} for a particular choice of parameters. The isocurvature constraint is much weaker here than in Fig.~\ref{fig2} because we have used, for illustration, a much smaller mass $m_{\rm s}$. Increasing $m_{\rm s}$ would tighten the limit. Also, adopting a smaller $m_\psi$ would reduce the abundance of $\psi$ produced via the mixing.

%
\section{Conclusions}
\label{conclusions}
%

We have performed a detailed analysis of the cosmological constraints on ultraweakly coupled Higgs portal scenarios with both bosonic and fermionic fields. By concentrating on dark matter production and isocurvature perturbations we have derived stringent constraints on couplings and masses in such models. Our most important results are the isocurvature bound on dark matter abundance, Eq.~\eqref{isocurvature_constraint}, and an absolute upper bound on dark matter particle mass, Eq.~\eqref{isocurvatureConstrainedMass}, calculated by making use of the isocurvature constraint and the total abundance of particles produced out from a primordial scalar field, Eq.~\eqref{highT_abundance}. 

The most interesting feature of the result~\eqref{isocurvatureConstrainedMass} is that it connects the inflationary energy scale $H_*$ to the portal sector parameters $m_{\rm DM}$ and $\lambda_{\rm s}$.  Qualitatively similar results were obtained 
already in \cite{Nurmi:2015ema}. Here we have extended the analysis accounting both for the decay of primordial condensate into portal sector fermions and for the standard freeze-in contribution to singlet dark matter. Our results show that although isocurvature bounds are alleviated in the limit of light fermions, they always strongly constrain the portal sector couplings and masses. We find that the standard freeze-in mechanism for dark matter production still remains viable, but in a limited parameter space which depends on the scale of inflation.

In conclusion, a detailed study of CMB anisotropies and primordial gravitational waves probing the inflationary scale may, in interesting and unexpected ways, shed light over different SM extensions. While we considered explicitly a particular model with a singlet fermion and a scalar field, our results are easily extended to constrain also other very weakly coupled dark matter models.

%
\section*{Acknowledgements}
%

We thank J. V\"aliviita for discussions. This work was financially supported by the Academy of Finland projects 257532  and 267842. TT is supported by the Research Foundation of the University of Helsinki and VV by Magnus Ehrnrooth foundation.

%
\appendix
\section{Decay rates of an oscillating background}
\label{append_decay_rates}
%

The singlet field $s$ remains nearly constant until it becomes effectively massive, when $V''\sim H^2$. After this it starts to oscillate with a decreasing envelope, which can be approximated analytically as follows in the quartic ($T \gsim T_{\rm tr}$) and quadratic ($T < T_{\rm tr}$) regimes:
\be
\label{s0_evolution}
\sigma_0(T) \simeq
\begin{cases}
3.7\times10^{-5} \left(H_*/10^{11}\mathrm{GeV}\right)^{1/2} \lambda_{\rm s}^{-3/8} g_*^{1/3} T \equiv \sigma_0^{\scriptscriptstyle (4)}(T), & T \gsim T_{\rm tr}\ ,\\
1.6\times10^{-7} \left(H_*/10^{11}\mathrm{GeV}\right)^{\scriptscriptstyle 3/4} \mu_{\rm s}^{-1/2} \lambda_{\rm s}^{-5/16} g_*^{1/2} T^{3/2} \equiv \sigma_0^{\scriptscriptstyle (2)}(T),     & T< T_{\rm tr}\ .
\end{cases}
\ee
The transition temperature is given by condition $\lambda_{\rm s} \sigma_0^{\scriptscriptstyle (4)}(T_{\rm tr})^2 = \mu_{\rm s}^2$. For temperatures above $T_{\rm tr}$ the singlet sees an effectively quartic potential $\lambda_{\rm s} s^4 \gg \mu_{\rm s}s^2$, whereas below $T_{\rm tr}$ the quadratic mass dominates.  In the quartic regime 
\be
s_0(t) = \sigma_0^{\scriptscriptstyle (4)}(t){\rm cn}(0.85 \lambda_{\rm s}^{1/2} \sigma_0^{\scriptscriptstyle (4)}(t) t,1/\sqrt{2}),
\ee 
where cn is the Jacobi cosine, and the oscillations can be divided into multiple tones, whereas in the quadratic regime 
\be
s_0(t) = \sigma_0^{\scriptscriptstyle (2)}(t) \cos(\mu_{\rm s}t),
\ee 
and the condensate oscillates with one frequency only.

The oscillating background generates an additional mass term for $s$, $h$ and $\psi$ particles, so that the masses are
\be
\label{adiabatic_masses}
\begin{aligned}
M_{\rm s}^2 &= \mu^2_{\rm s} + 3\lambda_{\rm s} s_0(t)^2, \\
M_{\rm h}^2 &= \mu_{\rm h}(T)^2 + \frac{\lambda_{\rm hs}}{2} s_0(t)^2, \\
M^2_{\psi} &= m^2_\psi + g^2 s_0(t)^2 ,
\end{aligned}
\ee
where $\mu_{\rm h}(T)^2\sim 0.1T^2$. At $T<T_\mathrm{EW}\approx 150\mathrm{GeV}$ the Higgs vacuum expectation value gives an additional contribution $\lambda_{\rm hs} v^2/2$ to $M_{\rm s}^2$ and $3\lambda_h v^2$ to $M_{\rm h}^2$. Note that the fermionic mass term is written in basis where it takes a real value, requiring a chiral rotation which transforms $g s\bar{\psi}\gamma_5\psi \to s\bar{\psi}(g_{\rm S}+ig_{\rm P}\gamma_5)\psi$ with $g_{\rm S}^2+g_{\rm P}^2 = g^2$.

An oscillating background field can lead to particle production~\cite{Abbott:1982hn, Ichikawa:2008ne,Nurmi:2015ema}. To derive the corresponding decay rates it is convenient to write the field and its square in Fourier series
\be
\begin{aligned}
\label{s_fourier}
s_0(t) &= \sum_{n=-\infty}^{\infty} \chi_n e^{+i\omega nt} , \\
s_0(t)^2 &= \sum_{n=-\infty}^{\infty}\zeta_n e^{-i2\omega nt} ,
\end{aligned}
\ee
where $\omega$ is the oscillation frequency of $s_0$.

The decay rate of the condensate energy density is given by
\be
\Gamma_{\rm s_0} = \frac{1}{\rho_{\rm s_0}} \sum_n \int \prod_{j=i,f}\frac{d^3p_j}{(2\pi)^3 2E_j} E_n |\mathcal{M}_n|^2 (2\pi)^4 \delta^4(p_n+{\textstyle \sum_i} p_i-{\textstyle \sum_f} p_f) \prod_i f_i \prod_f (1\pm f_f),
\ee
where
\be
\rho_{\rm s_0}(T) = 
\begin{cases}
\frac{\lambda_{\rm s}}{4}\sigma_0(T)^4, & T \gsim T_{\rm tr}\ ,\\
\frac{\mu_{\rm s}}{2}\sigma_0(T)^2,     & T< T_{\rm tr}\ ,
\end{cases}
\ee
is the energy density of the condensate, $E_n$ is the energy of the $n$th Fourier mode, $p_n = (E_n,0)$, $\mathcal{M}_n$ is the amplitude of the process $i\to f$ corresponding to $n$th Fourier mode, $f_j$ are phase space distribution functions, and $+$ applies for bosons and $-$ for fermions. 

Neglecting the blocking and stimulated emission factors, the decay rates of the condensate energy density induced by the interactions $\lambda_{\rm s} s_0(t)^2 s^2$, $i g s_0(t)\bar{\psi}\gamma_5\psi$ $\lambda_{\rm hs} s_0(t)^2 h^2$, $\lambda_{\rm s} s_0(t) s^3$ and $\lambda_{\rm hs} s_0(t) s h^2$ are, respectively, given by
\be
\begin{aligned}
\Gamma_{s_0\rightarrow ss} &= \frac{9\lambda_{\rm s}^2 \omega}{8\pi\rho_{\rm s_0}}\sum_{n=1}^{\infty}n|\zeta_n |^2\sqrt{1-\left(\frac{M_{\rm s}}{n\omega}\right)^2} ,  \\
\Gamma_{s_0\rightarrow \bar{\psi}\psi} &= \frac{\omega^3}{4\pi\rho_{\rm s_0}} \sum_{n=1}^{\infty}n^3|\chi_n |^2\left(g_S^2\left(1-\left(\frac{2 M_{\rm \psi }}{n\omega}\right)^2 \right)^{\frac{3}{2}}+g_P^2\sqrt{1-\left(\frac{2 M_{\rm \psi }}{n\omega}\right)^2} \right),  \\
\Gamma_{s_0\rightarrow hh} &=\frac{\lambda_{\rm hs}^{2}\omega}{8\pi\rho_{\rm s_0}}\sum_{n=1}^{\infty}n|\zeta_n |^2 \sqrt{1-\left(\frac{M_{\rm h}}{n\omega}\right)^2} ,  \\
\Gamma_{s\rightarrow ss} &= \frac{9\lambda_{\rm s}^2\omega}{8\pi\rho_{\rm s_0}}\sum_{n=1}^{\infty}\frac{n|\chi_n|^2}{M_{\rm s}}\sqrt{1-\left(\frac{2M_{\rm s}}{M_{\rm s}+n\omega} \right)^2}\frac{K_1(M_{\rm s}/T)}{K_2(M_{\rm s}/T)}n_{\rm s} , \\
\Gamma_{h\rightarrow sh} &= \frac{\lambda^2_{\rm hs}\omega}{2\pi \rho_{\rm s_0}}\sum_{n=1}^{\infty}n|\chi_n|^2 \frac{\sqrt{(n^2\omega^2 -M_{\rm s}^2 ) ((2 M_{\rm h} + n\omega)^2-M_{\rm s}^2)}}{2 (M_{\rm h} + n\omega)^2} \frac{K_1(M_{\rm h}/T)}{K_2(M_{\rm h}/T)}n_{\rm h} ,
\end{aligned}
\ee
where $K_j$ are the modified Bessel functions of the second kind and $n_{\rm j}$ denote particle number densities. The first three processes describe particle production from vacuum state, whereas the latter two are particle decays induced by the condensate. We have not written $\Gamma_{s_0\rightarrow sss}$ and $\Gamma_{s\rightarrow hh}$ because $s_0\rightarrow sss$ is negligible to $s_0\rightarrow ss$ and $s\rightarrow hh$ is insignificant because $M_{\rm s}\ll M_{\rm h}$. In contrast to~\cite{Ichikawa:2008ne} where particles were assumed to be massless, we have used adiabatic mass terms given by~\eqref{adiabatic_masses}.

Finally, we average the decay rates over one oscillation
\be
\left\langle \Gamma \right\rangle = \int_0^{\frac{2\pi}{\omega}} \frac{\mathrm{d}t}{2\pi} \Gamma(t).
\ee
In the quadratic potential only $s_0\to \bar{\psi}\psi$ is allowed, because $\omega = \mu_s$ and only $n=1$ mode is nonzero, whereas in the quartic potential $\omega = 0.85 \lambda_{\rm s}^{\scriptscriptstyle 1/2} \sigma_0^{\scriptscriptstyle (4)}(T)$ and $\chi_n\neq0$ for all odd $n$ and $\zeta_n\neq0$ for all $n$, so all processes except $s_0\to \bar{\psi}\psi$ (see Appendix \ref{fermi_statistics}) are allowed for some $n$.

%
\section{Constraints from fermion statistics}
\label{fermi_statistics}
%

The oscillating background produces fermions with a maximum momentum $p^2_{\rm F} = \omega^2 - m^2_{\psi}$ (see Appendix \ref{append_decay_rates} for details). By estimating that the produced fermions constitute degenerate Fermi gas, the maximum energy density of fermions produced from the condensate is
\be
\rho_{\rm F} = \frac{1}{8\pi^2} p_{\rm F}^4 .
\ee
Comparison to the energy density of the condensate, $\rho_{\rm s_0} \simeq V$, reveals that in the quartic regime
\be
\frac{\rho^{\scriptscriptstyle (4)}_{\rm F}}{\rho^{\scriptscriptstyle (4)}_{\rm s_0}} \simeq 10^{-2}\lambda_{\rm s} \ll 1 ,
\ee
i.e. fermion statistics renders the condensate decay into fermions inefficient.

In the quadratic regime
\be
\label{fermi_qdr}
\frac{\rho^{\scriptscriptstyle (2)}_{\rm F}}{\rho^{\scriptscriptstyle (2)}_{\rm s_0}} = \frac{1}{16\pi^2}\frac{m_{\rm s}^2}{\sigma_0(T)^2}\left(1- \left(\frac{2m_\psi}{m_{\rm s}}\right)^2 \right)^2 ,
\ee
where $\sigma_0$ is given by Eq.~\eqref{s0_evolution}. We have checked that the ratio~\eqref{fermi_qdr} becomes equal to one before photon decoupling in all cases under consideration. Therefore, the complete decay of the condensate by channel $s_0\to \bar{\psi}\psi$ is possible only in the quadratic regime.

%
\bibliography{freezein+fermions.bib}
%

\end{document}